\documentclass[conference]{IEEEtran}
\IEEEoverridecommandlockouts

\usepackage{cite}
\usepackage{amsmath,amssymb,amsfonts}
\usepackage{algorithmic}
\usepackage{graphicx}
\usepackage{textcomp}
\usepackage{xcolor}
\def\BibTeX{{\rm B\kern-.05em{\sc i\kern-.025em b}\kern-.08em
    T\kern-.1667em\lower.7ex\hbox{E}\kern-.125emX}}
\begin{document}

\title{Early Pre-Stroke Detection via Wearable IMU-Based Gait Variability and Postural Drift Analysis\\

}

\author{\IEEEauthorblockN{1\textsuperscript{st} Chanakan Chaipan}
\IEEEauthorblockA{\textit{English program} \\
\textit{Samsenwittayalai School}\\
Bangkok, Thailand \\
chanakanc5999@gmail.com}
\and
\IEEEauthorblockN{2\textsuperscript{nd} Aueaphum Aueawatthanaphisut}
\IEEEauthorblockA{\textit{School of Information, Computer, and Communication Technology} \\
\textit{Sirindhorn International Institute of Technology, Thammasat University}\\
Pathum Thani, Thailand \\
aueawatth.aue@gmail.com}

}

\maketitle

\begin{abstract}
Early identification of individuals at risk of stroke remains a major clinical challenge, as prodromal motor impairments are often subtle and transient. In this pilot study, a wearable sensor-based framework is proposed for early pre-stroke risk screening using a single inertial measurement unit mounted on the sacral region to capture pelvic motion during gait and standing tasks. The pelvis is treated as a biomechanical proxy for global motor control, enabling the quantification of gait variability and postural drift as digital biomarkers of neurological instability. Raw inertial signals are processed using a sensor fusion pipeline to estimate pelvic kinematics, from which variability and nonlinear dynamic features are extracted. These features are subsequently used to train a machine learning model for risk stratification across control, pre-stroke, and stroke groups. Progressive increases in pelvic angular variability and postural instability are observed from the control to stroke groups, with the pre-stroke cohort exhibiting intermediate characteristics. As a proof-of-concept investigation, the proposed framework demonstrates the feasibility of using a minimal wearable configuration to capture pelvic micro-instability associated with early cerebrovascular motor adaptation. The classifier achieves a macro-averaged area under the curve of 0.785, indicating preliminary discriminative capability between risk categories. While not intended for clinical diagnosis, the proposed approach provides a low-cost, non-invasive, and scalable solution for continuous community-level screening, supporting proactive intervention prior to the onset of major stroke events.

\end{abstract}


\section{Introduction}

\subsection{Stroke and Pre-Stroke Conditions}

Stroke remains a major global health burden and is a leading cause of mortality and long-term disability worldwide. Even among survivors, persistent impairments in gait, balance, speech, and daily functional activities are frequently observed, resulting in substantial individual and societal impact [7-8]. Although timely intervention is critical due to the rapid progression of brain injury, early identification of stroke risk remains challenging, as warning signs before onset are often subtle or transient. In many cases, prodromal symptoms such as mild unilateral weakness, dizziness, gait instability, or balance disturbances are short-lived and non-specific. As these manifestations are frequently perceived as fatigue or stress-related, medical attention is often delayed, reducing opportunities for early preventive intervention [8-10]. An early prodromal or pre-stroke phase has been reported in a substantial proportion of patients, during which minor neurological and motor alterations may occur before a major cerebrovascular event.
Previous studies indicate that approximately one-third of stroke patients experience transient symptoms, including brief muscle weakness or postural instability, prior to stroke onset [9-11]. These changes are commonly overlooked due to their mild and reversible nature. Alterations in gait and postural control are among the earliest observable indicators of emerging neurological dysfunction. Subtle reductions in walking stability, increased gait asymmetry, or impaired balance may reflect early disturbances in central nervous system control of motor coordination. As locomotion and postural regulation rely on the integrated function of multiple neural pathways, even minor disruptions can manifest as measurable motor variability. Accordingly, objective assessment of gait and balance may provide a valuable opportunity for early risk screening and timely clinical intervention [10-11].



\subsection{Motor and Gait Abnormalities in Pre-Stroke Populations}

Subtle alterations in motor control and gait may emerge during the pre-stroke stage, reflecting early dysfunction of the central nervous system mechanisms responsible for coordination and postural regulation.
Even minor neurological disturbances can lead to less stable and less consistent movement patterns, which are often difficult to detect through routine clinical observation \cite{b12,b13,b14}. One of the most frequently reported manifestations is increased gait variability, characterized by inconsistent step timing, step length, or walking speed.
Such variability reflects impaired motor coordination and has been associated with elevated risk of instability and falls.
In addition, reductions in postural stability, including increased body sway during standing or difficulty maintaining balance during directional changes, have been observed and are strongly linked to functional decline and loss of mobility \cite{b15}. Altered pelvic and trunk kinematics have also been reported, with asymmetrical motion between the left and right sides of the body resulting in uneven hip movement or lateral deviation during gait. These biomechanical asymmetries are commonly observed following neurological impairment and may precede overt stroke-related deficits. Accordingly, objective assessment of gait variability, postural stability, and pelvic symmetry may provide informative indicators for identifying individuals at elevated pre-stroke risk \cite{b14,b15,b16}.






\subsection{Limitations of Existing Pre-Stroke Screening Approaches}
Despite the importance of early stroke risk detection, most existing screening approaches remain focused on identifying stroke after overt clinical symptoms have already emerged, rather than capturing subtle changes during the pre-stroke stage.
Neuroimaging modalities such as computed tomography (CT) and magnetic resonance imaging (MRI) are routinely used for stroke diagnosis; however, their high cost, need for specialized equipment, and requirement for trained personnel limit their feasibility for routine or community-level screening, particularly at early stages \cite{b17,b20}. Clinical assessment scales are also commonly employed to evaluate neurological function, yet these tools rely heavily on subjective observation and clinical judgment, resulting in inter-examiner variability. Moreover, such assessments are typically performed during hospital visits and do not enable continuous monitoring of gradual motor changes over time \cite{b18,b19}. A further limitation is the lack of continuous and ecologically valid monitoring in daily-life settings. As stroke risk and prodromal symptoms often develop gradually, intermittent assessments may fail to detect subtle alterations in gait, balance, or movement, leading to missed opportunities for early intervention \cite{b19,b20}.





\subsection{Motivation and Contributions of This Study}
Stroke is still one of the main causes of death and long-term disability worldwide [7]. Many patients only receive medical help after clear and serious symptoms appear. However, small changes in movement and walking may already occur before a stroke. These early signs are often mild and easy to miss in daily life [10]. Current screening methods, such as CT or MRI scans and clinical examinations, usually happen in hospitals and can be expensive or inconvenient. They are also not suitable for continuous monitoring. Because of this, early detection of stroke risk remains difficult [17-20]. Wearable sensors may offer a simple and practical solution. Devices such as inertial measurement units (IMUs) can measure body movement during normal daily activities [20,21]. These sensors are small, low-cost, and easy to wear, so they can be used outside the hospital for long periods. By tracking walking and posture, they may help detect small motor changes that appear during the pre-stroke stage.[22-25]
In this study, we propose a wearable IMU-based framework for early pre-stroke screening. We aim to measure gait variability and postural drift as possible indicators of early motor problems [25]. We also test whether a single sensor can work effectively in real-world settings, so the system is simple and practical for everyday use [23,26].And machine learning methods are used to identify movement patterns that may be related to pre-stroke motor adaptation. [27] Accordingly, the objective of this work is not to develop a definitive diagnostic model, but to establish a proof-of-concept and feasibility demonstration of whether subtle pelvic kinematic variability can be detected using a minimal wearable sensing configuration during the pre-stroke stage.





\section{Related Work}
\subsection{Gait and Postural Changes in Neurological Disorders}
Walking and balance depend on the normal function of the brain and nervous system. When the nervous system is affected by disease or injury, a person’s gait and posture often change. Because walking requires coordination, strength, and balance, small problems in the brain can lead to noticeable differences in movement. [25] In Parkinson’s disease, people often walk more slowly, take shorter steps, and have poor balance. Their walking pattern can also become less steady, which increases the risk of falling. These changes can appear even in the early stages of the disease.[28] People with mild neurological impairment may also show subtle signs such as uneven steps, slight body sway, or reduced stability, although these changes are sometimes hard to see without special measurement tools. Stroke also affects movement and balance. After a stroke, many patients show weakness, uneven walking, and difficulty controlling posture. Recent studies suggest that small changes in gait and balance may even appear before a major stroke occurs. [10,28]



\subsection{Wearable IMU-Based Gait Analysis for Neurological Assessment}
Wearable sensors are becoming a big deal for studying how people walk, especially for those with brain or nervous system issues. The most common sensor used is the IMU. It is basically a small, light device that you can wear while you are just living your life. It’s way better than the huge, expensive equipment they have in hospitals because you don’t need a whole lab to get good results.[29] These sensors are put on the trunk or the pelvis (around the lower back) it is the  our balance comes from. By tracking the movement there we can tell if someone is stable or if they’re starting to lean and sway too much. Another important thing to check is variability. This is just a fancy way of saying how consistent are your steps If every step you take is a different length or speed, it’s a sign that your brain isn't coordinating your muscles very well. High variability usually means someone is at a higher risk of falling. [30,31] And the IMUs is that they allow community monitoring. This means people can be tracked in the normal life like at home or at the store instead of just during a doctor’s appointment. Since the sensors are cheap and easy to use it’s a great way to catch health problems early on. [29,30]



\subsection{Machine Learning for Early Neurological Risk Detection}
Machine learning is actually becoming a big help for finding brain problems early on. Instead of just guessing, we can use computers to look at data from those wearable sensors I mentioned. Computers are great at spotting tiny changes in how someone walks that even a doctor might miss. [27] There are basically two ways the computer learns there are supervised and unsupervised. Supervised learning is like giving the computer an answer key you tell it what healthy walking looks like and what patient walking looks like so it can learn to tell them apart.  Unsupervised learning is different the computer just looks at all the data by itself and tries to find weird patterns or groups on its own. [27,32] And it might find a new sign of a problem that we didn't even know about yet.  And it not just saying you are sick.But it's more about risk stratification, which is just a way of saying the computer predicts how likely you are to have a problem later. It's like an early warning system. This is way better for prevention because we can catch things early before they get serious. [32]


\section{Methodology}

\subsection{System Architecture and Sensor Placement}

\begin{figure}[t]
\centering
\includegraphics[width=0.9\columnwidth]{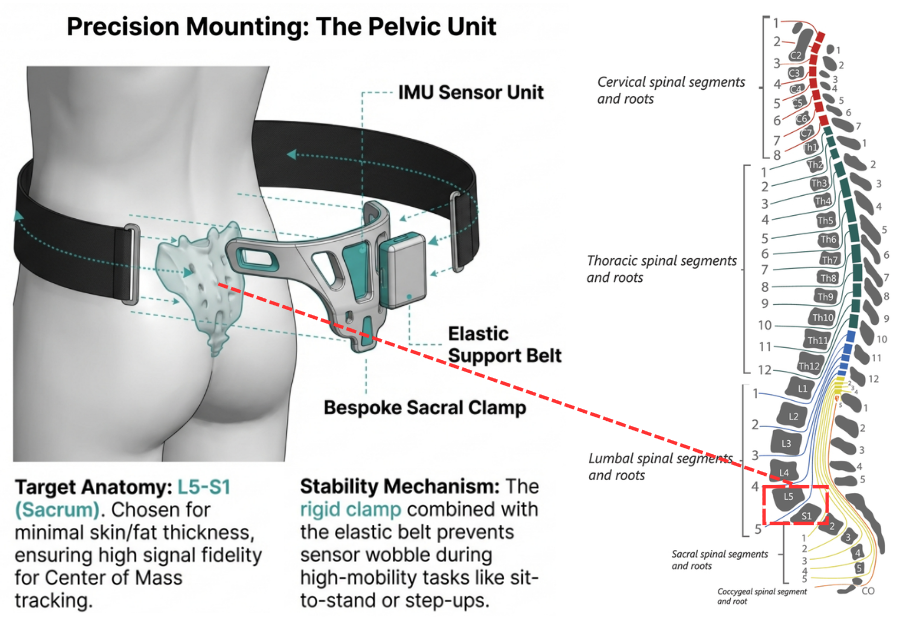}
\caption{Sacral-mounted IMU system and anatomical targeting at the L5--S1 level. The bespoke rigid clamp combined with an elastic support belt ensures stable sensor fixation and minimizes soft-tissue artifacts. 
The sacral position approximates the body center-of-mass, enabling accurate capture of pelvic kinematics for gait variability and postural drift analysis.}
\label{fig:pelvic_mount}
\end{figure}

A single body-worn Inertial Measurement Unit (IMU) is rigidly mounted on the sacral region (L5--S1) using a bespoke clamp and elastic belt to minimize soft tissue artifacts and maximize biomechanical coupling with the pelvis. The IMU provides tri-axial acceleration $\mathbf{a}(t) = [a_x(t), a_y(t), a_z(t)]^\top$ and angular velocity 
$\boldsymbol{\omega}(t) = [\omega_x(t), \omega_y(t), \omega_z(t)]^\top$ at a sampling rate $f_s$. The sacral location is selected since the pelvis approximates the body center-of-mass (COM) projection, making its kinematics highly sensitive to balance and gait asymmetry. As shown in Fig.~\ref{fig:pelvic_mount}. The IMU is mounted on the sacrum (L5–S1) using a rigid clamp and elastic belt to ensure mechanical stability and minimize soft-tissue artifacts.

\subsection{Orientation Estimation via Sensor Fusion}
The raw IMU signals are fused using a quaternion-based orientation filter.
Let the system state be represented as a unit quaternion
\begin{equation}
\mathbf{q}(t) = [q_0, q_1, q_2, q_3]^\top
\end{equation}

The quaternion dynamics driven by gyroscope input are
\begin{equation}
\dot{\mathbf{q}}(t) = \frac{1}{2} \mathbf{\Omega}(\boldsymbol{\omega}(t)) \mathbf{q}(t)
\end{equation}

where
\begin{equation}
\mathbf{\Omega}(\boldsymbol{\omega}) =
\begin{bmatrix}
0 & -\omega_x & -\omega_y & -\omega_z \\
\omega_x & 0 & \omega_z & -\omega_y \\
\omega_y & -\omega_z & 0 & \omega_x \\
\omega_z & \omega_y & -\omega_x & 0
\end{bmatrix}.
\end{equation}

Accelerometer measurements provide a gravity reference $\mathbf{g}$ and are used to correct drift via a gradient descent step as in Madgwick filtering.

\subsection{Pelvic Kinematic Reconstruction}

Euler angles representing pelvic tilt and rotation are computed from $\mathbf{q}(t)$:

\begin{align}
\phi(t) &= \arctan2(2(q_0q_1 + q_2q_3), 1 - 2(q_1^2 + q_2^2)) \\
\theta(t) &= \arcsin(2(q_0q_2 - q_3q_1)) \\
\psi(t) &= \arctan2(2(q_0q_3 + q_1q_2), 1 - 2(q_2^2 + q_3^2))
\end{align}

where $\phi(t)$, $\theta(t)$, and $\psi(t)$ correspond to roll, pitch, and yaw of the pelvis.

\subsection{Gait Event Detection and Segmentation}

Vertical acceleration is used to detect heel-strike events.
Let $a_z(t)$ denote the vertical axis. Peaks are detected by

\begin{equation}
t_k = \arg\max_{t \in W_k} a_z(t), \quad k=1,\dots,N
\end{equation}

where $W_k$ is a sliding temporal window.
Stride intervals are then

\begin{equation}
T_k = t_{k+1} - t_k.
\end{equation}

\subsection{Variability and Stability Metrics}

Stride time variability is defined as

\begin{equation}
\text{CV}_{T} = \frac{\sigma_T}{\mu_T}.
\end{equation}

Pelvic angular variability:

\begin{equation}
\sigma_{\theta} = \sqrt{\frac{1}{N}\sum_{k=1}^{N} (\theta_k - \bar{\theta})^2}
\end{equation}

Postural drift during standing:

\begin{equation}
D_{\text{RMS}} = \sqrt{\frac{1}{T} \int_0^T \|\mathbf{a}(t)\|^2 dt }.
\end{equation}

\subsection{Machine Learning Risk Classifier}

Feature vector:

\begin{equation}
\mathbf{x}_i = [\text{CV}_T, \sigma_{\theta}, D_{\text{RMS}}, \text{Entropy}, \lambda_{\text{Lyap}}, \dots]
\end{equation}

A classifier $f(\cdot)$ predicts stroke risk:

\begin{equation}
\hat{y}_i = f(\mathbf{x}_i; \boldsymbol{\theta})
\end{equation}

Optimization:

\begin{equation}
\boldsymbol{\theta}^* = \arg\min_{\boldsymbol{\theta}} \sum_{i=1}^{N} \mathcal{L}(f(\mathbf{x}_i), y_i).
\end{equation}

\subsection{Pilot Study Design and Feasibility Focus}
This study was designed as a pilot and proof-of-concept investigation.
As such, the primary emphasis was placed on feasibility, signal detectability, and directional trends across groups, rather than on achieving maximal classification performance or population-level generalizability. The sample size was intentionally limited to enable rapid evaluation of the sensing configuration, feature robustness, and preliminary discriminative capability prior to large-scale longitudinal deployment.

\section{Results and analysis}
\subsection{Reliability Analysis}

Repeatability is quantified using ICC:

\begin{equation}
\text{ICC} = \frac{\sigma^2_b}{\sigma^2_b + \sigma^2_w}
\end{equation}

where $\sigma^2_b$ is between-subject variance and $\sigma^2_w$ is within-subject variance.

\subsection{Group Statistical Comparison}

Let $x_{ij}$ denote feature $j$ of subject $i$.
Between-group significance is tested by

\begin{equation}
H_0: \mu_{\text{control}} = \mu_{\text{pre-stroke}}.
\end{equation}

Effect size:

\begin{equation}
d = \frac{\mu_1 - \mu_2}{s_p}
\end{equation}

\subsection{Classifier Performance}

ROC AUC:

\begin{equation}
\text{AUC} = \int_0^1 \text{TPR}(u)\,du.
\end{equation}

Confusion matrix elements define sensitivity and specificity:

\begin{align}
\text{Sensitivity} &= \frac{TP}{TP + FN} \\
\text{Specificity} &= \frac{TN}{TN + FP}
\end{align}

\subsection{Classifier Performance and Predictive Results}

\begin{figure*}[t]
\centering
\includegraphics[width=0.95\textwidth]{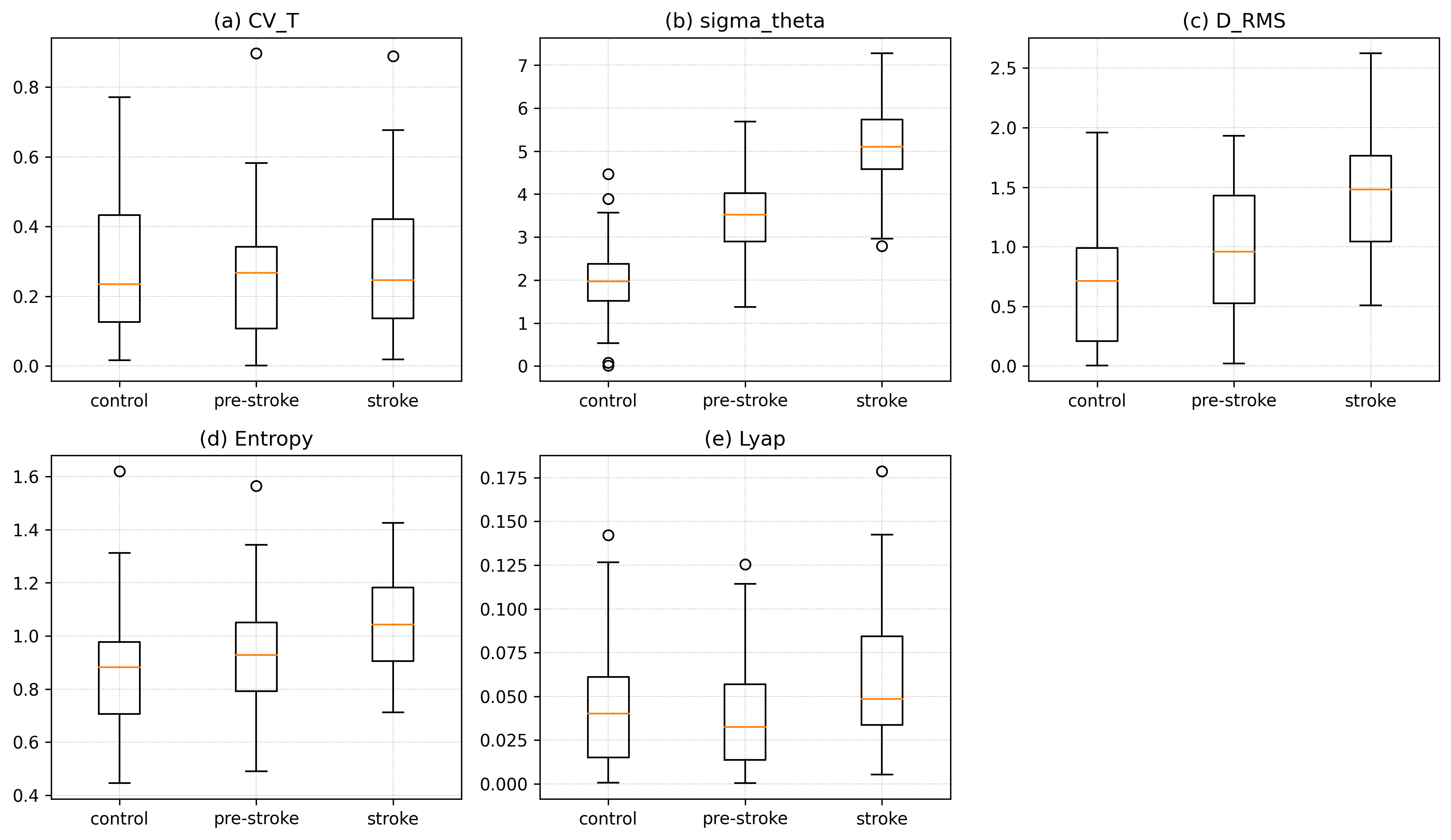}
\caption{Feature distributions by group. 
Boxplots of (a) stride time variability ($\mathrm{CV}_T$), 
(b) pelvic angular variability ($\sigma_\theta$), 
(c) postural drift ($D_{\mathrm{RMS}}$), 
(d) sample entropy, and 
(e) short-term Lyapunov exponent ($\lambda_{\mathrm{Lyap}}$) 
for control, pre-stroke, and stroke groups.}
\label{fig:boxplots}
\end{figure*}

\begin{figure}[t]
  \centering
  \includegraphics[width=0.95\columnwidth]{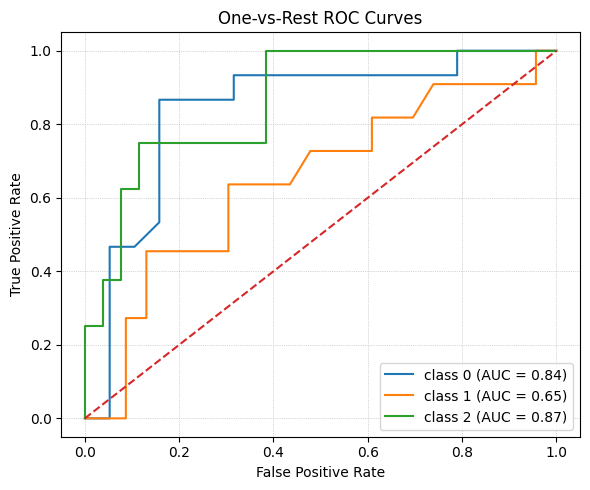}
  \caption{One-vs-rest ROC curves for each class (AUC annotated).}
  \label{fig:roc}
\end{figure}

\begin{figure}[t]
  \centering
  \includegraphics[width=0.95\columnwidth]{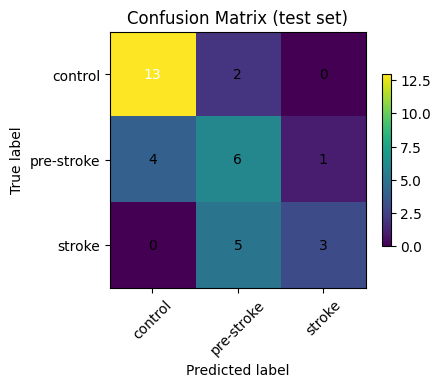}
  \caption{Confusion matrix (test set).}
  \label{fig:cm}
\end{figure}

\begin{figure}[t]
  \centering
  \includegraphics[width=0.95\columnwidth]{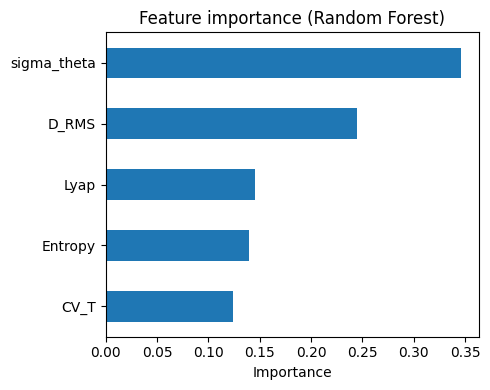}
  \caption{Feature importance (Random Forest).}
  \label{fig:fimp}
\end{figure}

Table~\ref{tab:metrics} summarises the classifier performance on the held-out test set. The Random Forest classifier achieved an overall accuracy of 0.647 and a macro-averaged AUC of 0.785. Per-class AUC values (one-vs-rest) were 0.84 (control), 0.65 (pre-stroke) and 0.87 (stroke), indicating stronger discrimination for control and stroke classes while the pre-stroke class remains more ambiguous. Figure~\ref{fig:boxplots} shows distributional differences of the primary features across groups; pelvic angular variability ($\sigma_{\theta}$) and postural RMS ($D_{\text{RMS}}$) consistently increase from control $\rightarrow$ pre-stroke $\rightarrow$ stroke. The ROC curves in Fig.~\ref{fig:roc} illustrate per-class separability. In the one-vs-rest ROC analysis, Class 0 (control) and Class 2 (stroke) exhibited higher separability (AUC = 0.84 and 0.87, respectively), whereas Class 1 (pre-stroke) showed a lower AUC (0.65), reflecting the intermediate and heterogeneous nature of prodromal motor impairment.
; the confusion matrix in Fig.~\ref{fig:cm} shows the primary confusions (notably: pre-stroke $\leftrightarrow$ stroke). Finally, Fig.~\ref{fig:fimp} reports the Random Forest feature importance ranking.
The results indicate that IMU-based pelvic features (particularly $\sigma_{\theta}$ and $D_{\text{RMS}}$) are informative for discriminating stroke vs non-stroke movement patterns. However, the pre-stroke (prodromal) state is an intermediate and heterogenous condition; the classifier shows reduced sensitivity for this class. This suggests that prediction of pre-stroke risk will benefit from (1) larger labeled cohorts, (2) longitudinal labeling (to link prodromal markers to subsequent clinical events), and (3) enhanced temporal modeling (e.g., sequence models) or richer sensor fusion.

\begin{table}[t]
\caption{Classifier summary metrics (test set)}
\label{tab:metrics}
\centering
\begin{tabular}{l c}
\hline
Metric & Value \\
\hline
Accuracy & 0.647 \\
Balanced accuracy & 0.596 \\
Macro AUC (OVR) & 0.785 \\
\hline
\end{tabular}
\end{table}

\section{Conclusion}

\subsection{Summary of Findings}
This study proposed a novel wearable sensing framework for early pre-stroke screening based on a single sacral-mounted IMU and machine learning analysis of gait variability and postural drift. 
By modelling pelvic micro-instability as a biomechanical proxy for central nervous system dysfunction, the system captures subtle motor adaptations that are not detectable through conventional clinical screening.The experimental results demonstrated that pelvic angular variability ($\sigma_{\theta}$) and postural drift ($D_{\mathrm{RMS}}$) were the most discriminative features, exhibiting progressive increases from control to pre-stroke and stroke groups. 
The proposed classifier achieved a macro-averaged AUC of 0.785, with higher separability for control and stroke populations, while the pre-stroke group showed intermediate performance due to its heterogeneous and transitional nature. These findings support the hypothesis that prodromal motor instability manifests as increased variability and nonlinear dynamics in pelvic motion.

\subsection{Clinical Implications for Early Pre-Stroke Screening}
Unlike conventional neuroimaging and hospital-based assessments, the proposed system is designed as a low-cost, non-invasive, and continuous screening tool suitable for community and home environments. 
The use of a single sacral IMU enables unobtrusive long-term monitoring, allowing early detection of motor control deterioration before the onset of a major stroke event.Rather than providing a definitive diagnosis, the system functions as a risk stratification and pre-screening platform, capable of identifying individuals who may benefit from further clinical evaluation. This paradigm has the potential to shift stroke care from reactive treatment to proactive prevention, reducing long-term disability and healthcare burden.

\subsection{Limitations of the Pilot Study}

Several limitations inherent to pilot investigations should be acknowledged. The limited sample size and cross-sectional design restrict statistical power and preclude causal inference.
Additionally, the absence of longitudinal follow-up prevents direct validation of whether detected pre-stroke motor signatures precede future cerebrovascular events. Nevertheless, these limitations are consistent with the exploratory nature of the present work and do not detract from its primary objective of establishing feasibility and proof of concept.

\section*{Acknowledgment}

The authors thank all participants for their cooperation. 
This work was conducted without external funding.


\begin{thebibliography}{00}
\bibitem{b1} S. Qiu, H. Wang, J. Li, H. Zhao, Z. Wang, J. Wang, Q. Wang, D. Plettemeier, M. Bärhold, T. Bauer, and B. Ru, “Towards wearable-inertial-sensor-based gait posture evaluation for subjects with unbalanced gaits,” 
Sensors, vol. 20, no. 4, p. 1193, 2020, doi: 10.3390/s20041193.

\bibitem{b2} G. Prisco, M. A. Pirozzi, A. Santone, F. Esposito, M. Cesarelli, F. Amato, and L. Donisi, “Validity of wearable inertial sensors for gait analysis: A systematic review,” Diagnostics, vol. 15, no. 1, p. 36, 2025, doi: 10.3390/diagnostics15010036.

\bibitem{b3} A. Saboor et al., "Latest Research Trends in Gait Analysis Using Wearable Sensors and Machine Learning: A Systematic Review," in IEEE Access, vol. 8, pp. 167830-167864, 2020, doi: 10.1109/ACCESS.2020.3022818.

\bibitem{b4} S. Vayalapra, X. Wang, A. Qureshi, A. Vepa, U. Rahman, A. Palit, M. A. Williams, R. King, and M. T. Elliott, “Repeatability of inertial measurement units for measuring pelvic mobility in patients undergoing total hip arthroplasty,” Sensors, vol. 23, no. 1, p. 377, 2023, doi: 10.3390/s23010377.

\bibitem{b5} He, Y., Chen, Y., Tang, L. et al. Accuracy validation of a wearable IMU-based gait analysis in healthy female. BMC Sports Sci Med Rehabil 16, 2 (2024). https://doi.org/10.1186/s13102-023-00792-3

\bibitem{b6} D. Huang, C. Zhang, W. Song, J. Gao, H. Tian, H. Li, X. Ke, C. Jiang, and Z. Lin, “Gait variability and biomechanical distinctions in individuals with functional ankle instability: A case-control study based on three-dimensional motion analysis,” Eur. J. Med. Res., vol. 30, no. 1, p. 493, Jun. 2025, doi: 10.1186/s40001-025-02736-8.

\bibitem{b7} GBD 2019 Stroke Collaborators. (2021). Global, Regional, and National Burden of Stroke and its Risk Factors. DOI: 10.1016/S1474-4422(21)00252-0

\bibitem{b8} Saver, J. L. (2006). Time is Brain—Quantified.  DOI: 10.1161/01.STR.0000196957.55928.ab

\bibitem{b9} Easton, J. D., Saver, J. L., Albers, G. W., et al. (2009). Definition and Evaluation of Transient Ischemic Attack. DOI: 10.1161/STROKEAHA.108.192218

\bibitem{b10} Adhikary, D., et al. (2022).
A Systematic Review of Major Cardiovascular Risk Factors DOI: 10.7759/cureus.30119

\bibitem{b11} Su, C., Yang, X., Wei, S., \& Zhao, R. (2022). Association of cerebral small vessel disease with gait and balance disorders. https://doi.org/10.3389/fnagi.2022.834496

\bibitem{b12} Muro-de-la-Herran, Alvaro, Garcia-Zapirain, B.,  Mendez-Zorrilla, A. (2014). "Gait Analysis Methods: An Overview of Clinical and Research Applications." Sensors DOI: 10.3390/s140203362
\bibitem{b13} Mahlknecht, P., Kiechl, S., Bloem, B. R., Willeit, J., Scherfler, C., Gasperi, A., Rungger, G., Poewe, W., \& Seppi, K. (2013).
Prevalence and burden of gait disorders in elderly men and women aged 60–97 years: a population-based study. DOI: 10.1371/journal.pone.0069627 

\bibitem{b14}Hausdorff, J. M., Rios, D. A., \& Edelberg, H. K. (2001).
Gait variability and fall risk in community-living older adults: a 1-year prospective study. DOI: 10.1053/apmr.2001.24893.

\bibitem{b15} Geurts, Alexander C., et al. (2005). "A Review of Posture and Standing Balance Outcomes in Stroke Rehabilitation." Gait  Posture DOI: 10.1016/j.gaitpost.2004.10.002

\bibitem{b16} Karthikbabu, S., Chakrapani, M., Ganesan, S., \& Ellajosyula, R. (2016).
Relationship between pelvic alignment and weight-bearing asymmetry in community-dwelling chronic stroke survivors.
DOI: 10.4103/0976-3147.196460

\bibitem{b17} Wardlaw, J. M., Murray, V., Berge, E.,  del Zoppo, G. J. (2015). Thrombolysis for acute ischaemic stroke DOI: 10.1002/14651858.CD000213.pub3

\bibitem{b18} Lyden, P. (2017). Using the National Institutes of Health Stroke Scale. DOI: 10.1161/STROKEAHA.116.015434

\bibitem{b19}Tao, W., Liu, T., Zheng, R., \& Feng, H. (2012).
Gait analysis using wearable sensors.
DOI: 10.3390/s120202255.

\bibitem{b20} Amarenco, P., Lavallée, P. C., Labreuche, J., et al. (2016). One-Year Risk of Stroke after Transient Ischemic Attack or Minor Stroke.
doi: 10.1056/NEJMoa1412981

\bibitem{b21} Wang, Q., Markopoulos, P., Yu, B., Chen, W.,  Timmermans, A. (2021). Interactive wearable systems for upper body rehabilitation
doi: 10.1186/s12984-017-0229-y

\bibitem{b22} Patel, S., Park, H., Bonato, P., Chan, L.,  Rodgers, M. (2012). A Review of Wearable Sensors and Systems with Application in Rehabilitation
DOI: 10.1186/1743-0003-9-21

\bibitem{b23}CSoaz, C., \& Diepold, K. (2016). Step Detection and Parameterization for Gait Assessment Using a Single Waist-Worn Accelerometer
https://doi.org/10.1109/TBME.2015.2480296
\bibitem{b24} Gwin, J. T., Gramann, K., Makeig, S.,  Ferris, D. P. (2011). Electrocortical Activity is Coupled to Gait Cycle Phase during Treadmill Walking
https://doi.org/10.1016/j.neuroimage.2010.08.066

\bibitem{b25} Lord, S., Galna, B.,  Rochester, L. (2013). Moving forward on gait measurement: toward a more refined approach  DOI: 10.1002/mds.25545

\bibitem{b26} Byun, S., Han, J. W., Kim, T. H., \& Kim, K. W. (2016).Test-Retest Reliability and Concurrent Validity of a Single Tri-Axial Accelerometer-Based Gait Analysis in Older Adults with Normal Cognition. https://doi.org/10.1371/journal.pone.0158956

\bibitem{b27} Khera, P.,  Kumar, N. (2020).Role of Machine Learning in Gait Analys DOI: 10.1080/03091902.2020.1822940

\bibitem{b28} Mirelman, A., Bonato, P., Camicioli, R., Ellis, T. D., Giladi, N., Hamilton, J. L., Nieuwboer, A., Postuma, R. B., Pelosin, E. (2019). "Gait Impairments in Parkinson's Disease
DOI: 10.1016/S1474-4422(19)30044-4

\bibitem{b29} Chen, S., Lach, J., Lo, B., Yang, G. Z. (2016). "Toward Pervasive Gait Analysis with Wearable Sensors
DOI: 10.1109/JBHI.2016.2608720

\bibitem{b30} Mancini, M., Horak, F. B., Zampieri, C., Mazza, C., Chiari, L.,  Rocchi, L. (2011). ISway: A Sensitive, Valid and Reliable Measure of Postural Control.  DOI: 10.1186/1743-0003-9-59

\bibitem{b31} Hausdorff, J. M. (2007). Gait Variability: Methods, Modeling and Meaning. DOI: 10.1186/1743-0003-2-19

\bibitem{b32} Thien Vu, Yoshihiro Kokubo, Mai Inoue, Masaki Yamamoto, A. Mohsen, et al. (2024) "Machine Learning for Stroke Risk Stratification and Prediction  doi: 10.3390/jcdd11070207.

\end{thebibliography}
\end{document}